\documentclass[12]{article}

\def\ba{\mathbf{a}}   %\def\ba{{\bf a}}
\def\bb{\mathbf{b}}   %\def\bb{{\bf b}}
\def\be{\mathbf{e}}   %\def\bd{{\bf e}}
\def\bbf{\mathbf{f}}   %\def\bff{{\bf f}}

\def\C{\mathbb{C}} %\def\C{I\!\!\!C} %\def\C{{C\kern-.647em I}}
\def\G{\mathbb{G}} % \def\G{I\!\!\!G}

\def\R{\mathbb{R}}  %\def\R{I\!\!R}

\def\no{\noindent}

\def\beq{\begin{equation}}
\def\eeq{\end{equation}}

\def\w{\wedge}

\usepackage{array}
\usepackage{tabularx}
\usepackage{amsfonts}
\usepackage{amssymb}
\usepackage{tikz-cd}
\usetikzlibrary{matrix,arrows,decorations.pathmorphing}

\def\bpm{\begin{pmatrix}}
	\def\epm{\end{pmatrix}}
\def\w{\wedge}

\begin{document}
\author{Garret Sobczyk\\ Departamento de Actuaría F\'isica y Matem\'aticas,\\Universidad de las Am\'ericas-Puebla,\\72820 Puebla, Pue., M\'exico\\
	garretudla@gmail.com\\}
\title{Itinerant Quantum Integers: The Language of Quantum Computers} 
\maketitle
\begin{abstract}
	The concept of positively and negatively compatible null vectors arises in the study of Clifford geometric algebras with a Lorentz-Minkowski metric. In 
	previous works, the basic properties of such algebras have been set down in terms of
	 a new principle of {\it quantum duality}.
	In the present work, the same structure is studied in terms of real and complex quantum integers, which generalize the real and complex number systems. It seems natural to identify a {\it qubit} as a pair of compatible null vectors; the {\it up state} of the qubit being their {\it sum}, and the {\it down state} being their {\it difference}. Basic identities are developed to make calculations routine, and two different representations of the symmetric group are given. 
	
	\smallskip
\no {\em AMS Subject Classification:} 05C20, 15A66, 15A75
\smallskip

\no {\em Keywords:} Clifford geometric algebra,  Grassmann algebra, Lorentzian spacetime, quantum duality. 

\end{abstract}

\section*{0 \ \ Introduction}

How the concept of {\it compatible null vectors}, which I hereby dub {\it quantum duality}, reflects in many ways the standard definition of {\it dual vector spaces} in linear algebra, has been explored in previous works by the author \cite{ICAtalk,GS2022}. The purpose of this work is to introduce {\it real} and {\it complex} {\it quantum integers}, and simple algebraic rules that makes calculations in these unusual algebras routine. 

Section 1, reviews the  properties of the closely related symmetric algebras of positively and negatively compatible null vectors, which make up the generators of the geometric algebras $\G_{1,n}$ and $\G_{n,1}$, respectively, \cite{SobHAL2023}. The idea that linear algebra could equally well have been developed from Grassmann's algebra, together with the quite different Multiplication Table of compatible null vectors over the complex numbers, is explored.

Section 2, formally introduces the concept of a real and imaginary quantum integers, before studying in detail properties of real quantum numbers. Real quantum integers are an algebra ${\cal A}_{n+1}^+$ of positively compatible null vectors, whereas purely imaginary quantum integers are equivalent to an algebra ${\cal A}_{n+1}^-$ of negatively compatible null vectors \cite{Sob2023}. Complex quantum integers are positively compatible quantum integers taken over the complex numbers. Various identities are presented for the real quantum integers that help make calculations routine. 

Section 3, discusses how complex quantum numbers are related to real and complex matrix algebras, combinatorial mathematics and quantum mechanics. It seems natural to identify a {\it qubit} as a pair of compatible null vectors; the {\it up state} of the qubit being their {\it sum}, and the {\it down state} being their {\it difference}. There is much work to be done in exploring the relationship of this work to different approaches found in the literature \cite{TimD2001,KH2017,kauff2022,kauff2021}. 
 
 \section{The geometric algebras of postively and negatively compatible null vectors}

\begin{itemize}
	
	\item {\it Nilpotents} are algebraic quantities $x \ne 0$ with the property that $x^2=0$. They are added together using the same rules for the {\it addition and multiplication by scalars ${\cal F}$}, as the real or complex numbers. The trivial nilpotent is denoted by  $0$.
	\item A set of nilpotents 		
		${\cal A}_n:=\{\ba_1, \ldots , \ba_n\}_{\cal F}$ is said to be {\it multiplicatively uncorrelated} over ${\cal F}$, if 
	 for all $\ba_i,\ba_j \in {\cal A}_n$,
	 \beq \ba_i \ba_j + \ba_j \ba_i =0.\label{uncorrel} \eeq
	A set of uncorrelated nilpotents over a field $\cal F$ are called {\it null vectors}, and generate a {\it Grassmann algebra} ${\cal G}_n({\cal F})$, provided they are {\it linearly independent over} ${\cal F}$, satisfying
	 \beq \ba_1 \w \cdots \w \ba_n \not= 0,\label{Gcorrelated}  \eeq 
	 \cite{Dieu}.
	 More general fields ${\cal F}$ can be considered as long as {\it characteristic} ${\cal F}\not=2$. 
		
\item A set ${\cal A}_{n+1}^\pm:=\{\ba_1, \ldots, \ba_{n+1}\}$ of $n+1$ null vectors are said to be {\it positively} or {\it negatively correlated} if they satisfy the extra {\it quantum duality} condition that
\beq  2\ba_i \cdot \ba_j:=\ba_i \ba_j + \ba_j \ba_i =\pm(1-\delta_{ij}),  \label{alg-dual}  \eeq
respectively, with the {\it inner} and {\it outer products} denoted by $\ba_i \cdot \ba_j$ and 
\beq 2\ba_i \w \ba_j :=\ba_i\ba_j - \ba_j \ba_i, \label{outerp}  \eeq respectively.
They generate the $2^{n+1}$-dimensional {\it Clifford geometric algebras} $\G_{1,n}$ and $\G_{n,1}$, respectively \cite{WKing,SobHAL2023,Sob2023}.
\end{itemize}

\begin{table}[h!]
\begin{center}
	\caption{PC Multiplication table.}
	\label{table3.1}
	\begin{tabular}{c|c|c|c|c} 
		& $\ba_i $  & $ \ba_j $  & $\ba_i \ba_j $  & $ \ba_j \ba_i $ \cr 
		\hline
		$ \ba_i$  &   0   & $ \ba_i \ba_j $     & $  0 $         & $ \ba_i $ \cr 
		\hline         
		$ \ba_j $  &  $ \ba_j \ba_i $  & $ 0 $     & $  \ba_j $         & $ 0 $ \cr 
		\hline     
		$ \ba_i \ba_j $    &  $\ba_i $    &  0  &  $\ba_i \ba_j $    & 0  \cr
		\hline
		$ \ba_j \ba_i $  &  0   &  $\ba_j $   &  0     & $\ba_j \ba_i$   \cr
	\end{tabular}
\end{center}
\end{table}

Given above is the multiplication table for a set of positively  correlated null vectors $\ba_i , \ba_j$, for $1 \le i<j \le n+1$, define
\[ A_k := \sum_{i=1}^k \ba_i.  \] 
The geometric algebra 
\[ \G_{1,n}:=\R(\be_1,\bbf_1,\ldots, \bbf_n) ,\] 
where $\{ \be_1, \bbf_1, \ldots \bbf_n \}$ is the {\it standard basis} of anticommuting orthonormal vectors, with $\be_1^2=1$ and $\bbf_1^2 = \cdots = \bbf_n^2 =-1$, \cite[p.71]{Sob2019}.
Alternatively, the geometric algebra $\G_{1,n}$ can be defined by
\[ \G_{1,n}:=\R(\ba_1,\ldots, \ba_{n+1})={\cal A}_{n+1}^+,   \]
where   
$\{\ba_1, \ldots, \ba_{n+1} \}$ is a set of positively compatible null vectors satisfying the multiplication Table 1. In this case, the standard basis vectors of $\G_{1,n}$ can be defined by $\be_1=\ba_1+\ba_2=A_2$, $\bbf_1=\ba_1-\ba_2=A_1- \ba_2 $, and for $2\le k\le n$
\beq \bbf_k=\alpha_k \Big(A_k-(k-1)\ba_{k+1}\Big),   \label{precursive} \eeq  
where $\alpha_k := \frac{-\sqrt 2}{\sqrt{k(k-1)}}$. 

If we define a second set $\{\bb_1, \ldots, \bb_{n+1}\}_\R$, by $\bb_j = i \ba_j$ for
$j=1, \ldots, j_{n+1}$, and assume that they are negatively compatible satisfying (\ref{alg-dual}), then, over the real numbers, they generate the geometric algebra ${\cal A}_{n+1}^-\equiv  \G_{n,1}$, \cite{SobHAL2023}. However, care must be taken when considering complex null vectors of the form
$\ba_i+i \ba_j=\ba_i+\bb_j\in {\cal A}_{n+1}^+(\C)$, because bivectors will be real, and trivectors imaginary.
Also, 
\[ (\ba_j+i\ba_k)^2 = 2 i\ba_j\cdot \ba_k=i.  \]
 Different interesting geometric interpretations arise when considering ${\cal A}_{n+1}^+(\C)$, ${\cal A}_{n+1}^-(\C)$, and ${\cal A}_{n+1}^\pm(\R)$. 
 Such issues are further discussed in the last section of this work.

\subsection{Change of basis formulas for $\G_{1,n} \equiv {\cal A}_{n+1}^+(\R)$} 
For the remainder of this work, vectors and null vectors are no longer be denoted in boldface. 

\[ \pmatrix{a_1 \cr a_2 \cr \cdot \cr \cdot \cr  a_8} =T_8 \pmatrix{e_1 \cr f_1 \cr \cdot \cr \cdot \cr  f_7}  \]
for

\[ T_8=  \pmatrix{\frac{1}{2} &\frac{1}{2}  & 0 &0 &0&0 &0 &0  \cr 
	\frac{1}{2} &-\frac{1}{2}  & 0 &0 &0&0 &0 &0 \cr 1 & 0  & 1 &0 &0&0 &0 &0 \cr 1 & 0  & \frac{1}{2} &\frac{\sqrt 3}{2} &0&0 &0 &0 \cr
	1 & 0  & \frac{1}{2} &\frac{1}{2 \sqrt 3}&\sqrt{\frac{ 2}{3}} &0 &0&0  
	\cr  1 & 0  & \frac{1}{2} &\frac{1}{2 \sqrt 3}&\frac{ 1}{2\sqrt{6}} &\frac{\sqrt 5}{2\sqrt 2} &0&0 
	\cr  1 & 0  & \frac{1}{2} &\frac{1}{2 \sqrt 3}&\frac{ 1}{2\sqrt{6}} &\frac{1}{2\sqrt 10} &\sqrt{\frac{ 3}{ 5}}&0 
	\cr   1 & 0  & \frac{1}{2} &\frac{1}{2 \sqrt 3}&\frac{ 1}{2\sqrt{6}} &\frac{1}{2\sqrt 10} &\frac{ 1}{2\sqrt{15}}&\frac{\sqrt 7}{2\sqrt 3} }, \]
and

\[\pmatrix{e_1 \cr f_1 \cr \cdot \cr \cdot \cr  f_7}  =T_8^{-1} \pmatrix{a_1 \cr a_2 \cr \cdot \cr \cdot \cr  a_8},  \]
for
\[ T_8^{-1}=  \pmatrix{1 &1  & 0 &0 &0&0 &0 &0  \cr 
	1 &-1  & 0 &0 &0&0 &0 &0 
	\cr -1 & -1  & 1 &0 &0&0 &0 &0 
	\cr -{\frac{ 1}{\sqrt 3}} &-{\frac{ 1}{\sqrt 3}}  &-{\frac{ 1}{\sqrt 3}}&{\frac{ 2}{\sqrt 3}} &0&0 &0 &0
	\cr
	-{\frac{ 1}{\sqrt 6}} &-{\frac{ 1}{\sqrt 6}}  &-{\frac{ 1}{\sqrt 6}}&-{\frac{ 1}{\sqrt 6}} &\sqrt \frac{3}{2}&0 &0 &0
	\cr   -{\frac{ 1}{\sqrt 10}} &-{\frac{ 1}{\sqrt 10}}  &-{\frac{ 1}{\sqrt 10}}&-{\frac{ 1}{\sqrt 10}} &-\frac{1}{\sqrt 10}&2 \sqrt \frac{2}{5} &0 &0
	\cr -{\frac{ 1}{\sqrt 15}} &-{\frac{ 1}{\sqrt 15}}  &-{\frac{ 1}{\sqrt 15}}&-{\frac{ 1}{\sqrt 15}} &-\frac{1}{\sqrt 15}&-\frac{1}{\sqrt 15} &\sqrt \frac{5}{3} &0
	\cr    -{\frac{ 1}{\sqrt 21}} &-{\frac{ 1}{\sqrt 21}}  &-{\frac{ 1}{\sqrt 21}}&-{\frac{ 1}{\sqrt 21}} &- \frac{1}{\sqrt 21}&- \frac{1}{\sqrt 21} &- \frac{1}{\sqrt 21} &2 \sqrt \frac{3}{7} }, \]
\cite{Sob2023}.

	Given any two distinct $a_i,a_j \in {\cal A}_{n+1}^+$, let $e:=a_{i}+a_{j}$ and $f:=a_{i}-a_{j}$.
	It easily follows from (\ref{alg-dual}) that
	\beq e^2 =(a_i+a_j)(a_i+a_j)=a_i^2+(a_ia_j+a_ja_i)+ a_j^2  =1,\label{plusaij} \eeq
and 
	\beq f^2 =(a_i-a_j)(a_i-a_j)=a_i^2-(a_ia_j+a_ja_i)+ a_j^2  =-1.\label{minusaij} \eeq
Furthermore, the orthonormal vectors $e$ and $f$ are {\it anticommutative}, defining the {\it bivector} 
\beq ef = (a_i+a_j)(a_i-a_j)=a_j a_i-a_i a_j=-(a_i a_j-a_j a_i) = -fe. \label{anticommutef} \eeq
Once the {\it seed null vectors} $a_i$ and $a_j$ are chosen determining $e$ and $f$, the other null vectors $a_k$ used to define the successive vectors $f_k$'s can randomly be chosen using the recursive definition (\ref{precursive}).		

%\subsection{Basic identities in ${\cal A}_{n+1}^+$}
The {\it canonical basis} of the algebra ${\cal A}_{n+1}^+$ over the real numbers consists of elements of the form
\beq {\cal A}_{n+1}^+:=\{ 1, a_1, \ldots, a_{n+1}, \ldots, a_{\lambda_1 \cdots \lambda_k}, \ldots, a_{1\cdots n+1}\}_\R, \label{canonicalbasis} \eeq 
where  $1\le \lambda_1 <  \cdots < \lambda_{k}\le n+1$. Since there are
$\pmatrix{n+1 \cr k}$ elements 
\[a_{\lambda_1 \cdots \lambda_k}:=a_{\lambda_1}\cdots a_{\lambda_k}   \]
of {\it order} $k$, there are a total of
$2^{n+1}$ elements in the canonical basis of
${\cal A}_{n+1}^+$.

\section{Quantum Numbers}
In the definition of the canonical basis (\ref{canonicalbasis}) of ${\cal A}_{n+1}^+$, the basis elements are completely defined in terms of sequences of ordered positive integers, 
\[ 1\le \lambda_1 <\cdots <\lambda_k \le n+1.\]
This suggests that a more efficient notation can be employed. To keep matters as simple as possible, let us restrict our attention to the case when $n+1 \le 6$ for $n=5$. For this case, we express the equivalent null vector generators of the  canonical basis (\ref{canonicalbasis}) for ${\cal A}_6^+$, by 
\beq {\cal A}_6^+ :=gen\{ \hat 1 , \ldots, \hat 6   \}_\R . \label{canbasis6} \eeq

For example, the full canonical basis for $n+1=3$, over the real numbers,  is
\beq {\cal A}_3^+ := \{1, \hat 1, \hat 2, \hat 3, \hat 1 \hat 2, \hat 1 \hat 3, \hat 2 \hat 3, \hat 1 \hat 2 \hat 3     \}_\R. \label{canonical3} \eeq
The {\it order} of a quantum integer in ${\cal A}_3^+$ is $0,1,2,3$. By definition, the {\it order} of a quantum number $x$ is zero iff $ x \in \R$. The quantum number
$\hat v = \hat 1 \hat 3\in {\cal A}_3^+$ is of order $2$ since it is the product of the quantum null vectors $\hat 1,\hat 3$, etc., etc.. The concept of {\it order} is defferent than the concept of the {\it grade} of a geometric number. For example, whereas the order of the quantum number $\hat 1\hat{2}\hat{3}$ is $3$, the geometric algebra identity
\beq \hat1\hat2\hat{3}=\hat 1(\hat2\cdot\hat3+\hat 2 \w \hat 3)=\hat 1(\frac{1}{2}+\hat 2\w \hat3)=\frac{1}{2}(\hat 1-\hat 2+\hat 3)-
\hat 1\w 
\hat 2\w \hat 3,  \label{gaident123} \eeq
shows that the element $\hat1\hat2\hat3$ is of mixed grades $1$ and $3$. Also, the order of a general quantum number $\hat x\in {\cal A}_{n+1}^+$ is $k$, iff it is the sum of a linear combination of elements of order $k$ in the canonical basis of ${\cal A}_{n+1}^+$, otherwise it is of {\it mixed order}.
 
 Given below is a Multiplication Table for the canonical basis elements of ${\cal A}_3^+$. 

\[   \pmatrix{1 & \hat 1  & \hat 2 & \hat 3 & \hat1\hat 2 &\hat1\hat 3  &\hat2\hat 3  &\hat 1\hat2\hat3  \cr 
\hat1 & 0  & \hat1\hat 2 & \hat1\hat 3 & 0 & 0  &\hat1\hat2\hat 3  & 0  \cr 
\hat2 &1- \hat1\hat2  &0 & \hat2\hat 3 & \hat 2 &\hat3-\hat1\hat2\hat 3  & 0  &\hat2\hat3  \cr
\hat3 &1-\hat1 \hat3  &1-\hat 2 \hat3 & 0 & \hat 2-\hat1+\hat1\hat 2\hat3 &\hat 3  &\hat 3  & \hat1\hat2-\hat1\hat3  \cr 
\hat1 \hat2 & \hat1  & 0 & \hat1\hat2\hat 3 & \hat1\hat 2 &\hat1\hat 3  & 0  &\hat 1\hat2\hat3  \cr
 \hat1\hat3 & \hat1  & \hat 1 -\hat 1 \hat2\hat3  & 0 & \hat1\hat 2 &\hat1\hat 3  &\hat1\hat 3  &\hat1\hat2\hat3  \cr 
 \hat2\hat3 & \hat2-\hat3+\hat1\hat2\hat3  & \hat 2 & 0 &-1+ \hat2\hat3  +\hat{1}\hat2 &\hat2\hat 3  &\hat2\hat 3  & -\hat3+\hat1\hat2\hat3  \cr
 \hat1\hat2\hat3 & \hat1\hat2-\hat1\hat3  & \hat1\hat 2 & 0 & \hat1\hat 2\hat3-\hat 1 &\hat1\hat2\hat 3  &\hat1\hat2\hat 3  &-\hat 1\hat3   }. \]

The key to reducing the product of two quantum numbers $\hat u$ and $\hat v \in {\cal A}_{n+1}^+$ to canonical form is to develop identities that allow the elimination of repetitions that might occur in the product.
For example, in the Multiplication Table given for ${\cal A}_3^+$, using associativity and (\ref{alg-dual}), we calculated
\[ (\hat 1\hat2\hat3)\hat 1=\hat1\hat2(1-\hat1\hat3)=\hat1\hat2 - \hat1\hat2\hat1\hat3=\hat1\hat2-\hat1 \hat3. \]
From this identity, we successively calculated
 \beq (\hat 1\hat2\hat3)\hat 1\hat 2=-\hat1\hat3\hat2 =-\hat1(1-\hat2\hat3)=-\hat1+\hat1\hat2\hat3 ,  \label{square123}  \eeq
and \[(\hat1\hat2\hat3)^2=(-\hat1+\hat1\hat2\hat3)\hat3= -\hat1\hat3.   \] 

Higher order identities, like (\ref{square123}), up to $n+1=6$, will now be similarly calculated. The proofs of general formulas for the products of null vectors is by induction, but will not be given here. 
For $n+1=4$,
\beq (\hat1\hat2\hat3\hat4)\hat1 =\hat1\hat2\hat3- \hat1\hat2\hat4+\hat1\hat3\hat4 , \label{hat12341} \eeq 
\[  (\hat1\hat2\hat3\hat4)\hat1\hat2 =\hat1\hat3 - \hat1\hat4+\hat1\hat2\hat3\hat4, \ \ (\hat1\hat2\hat3\hat4)\hat1\hat2\hat3 = - \hat1+\hat1\hat3\hat4+\hat1\hat2\hat3,  \]
and
\[ (\hat1\hat2\hat3\hat4)^2 = -\hat1 \hat4+\hat1\hat2\hat3\hat4 .  \]

For $n+1=5$,
\beq (\hat1\hat2\hat3\hat4\hat5)\hat1 =\hat1\hat2\hat3\hat4- \hat1\hat2\hat3\hat5+\hat1\hat2\hat4\hat5-\hat1\hat3\hat4\hat5 , \label{hat123451} \eeq 
\[  (\hat1\hat2\hat3\hat4\hat5)\hat1\hat2 =\hat1\hat2\hat3\hat4\hat2 - \hat1\hat2\hat3\hat5\hat2+\hat1\hat2\hat4\hat5\hat2-\hat1\hat3\hat4\hat5\hat2=-\hat1\hat3\hat4\hat5\hat2,  \]
\[ (\hat1\hat2\hat3\hat4\hat5)\hat1\hat2\hat3 = -\hat1\hat3\hat4\hat5\hat2\hat3= -\hat1\hat3\hat4\hat5 +\hat1\hat3\hat4\hat2-\hat1\hat3\hat5\hat2+\hat1\hat4\hat5\hat2,   \]
\[ (\hat1\hat2\hat3\hat4\hat5)\hat1\hat2\hat3\hat4 = -\hat1\hat3\hat4\hat5 +\hat1\hat3\hat4\hat2-\hat1\hat3\hat5\hat2+\hat1\hat4\hat5\hat2 = -\hat1\hat4\hat2+\hat1\hat4\hat5-\hat1\hat3\hat5\hat2\hat4,\]
and
\[(\hat{1}\hat{2}\hat{3}\hat{4}\hat{5})^2=-\hat1\hat2\hat3\hat5+\hat1\hat2\hat4\hat5-\hat1\hat3\hat4\hat5.  \]

For $n+1=6$,
\beq (\hat1\hat2\hat3\hat4\hat5\hat6)\hat1 =\hat1\hat2\hat3\hat4\hat5- \hat1\hat2\hat3\hat4\hat6+\hat1\hat2\hat3\hat5\hat6-\hat1\hat2\hat4\hat5\hat6+\hat1\hat3\hat4\hat5\hat6 , \label{hat1234561} \eeq 
\[  (\hat1\hat2\hat3\hat4\hat5\hat6)\hat1\hat2 =\hat1\hat2\hat3\hat4\hat5\hat6,  \]
\[ (\hat1\hat2\hat3\hat4\hat5\hat6)\hat1\hat2\hat3 = \hat1\hat2\hat3\hat4\hat5 -\hat1\hat2\hat3\hat4\hat6 +\hat1\hat2\hat3\hat5\hat6-\hat1\hat2\hat4\hat5\hat6+\hat1\hat3\hat4\hat5\hat6,   \]
\[ (\hat1\hat2\hat3\hat4\hat5\hat6)\hat1\hat2\hat3\hat4 = \hat1\hat2\hat3\hat5 -\hat1\hat2\hat3\hat5\hat4\hat6-\hat1\hat2\hat4\hat5+\hat1\hat2\hat4\hat6 +\hat1\hat3\hat4\hat5-\hat1\hat3\hat4\hat6,\]
\[(\hat{1}\hat{2}\hat{3}\hat{4}\hat{5}\hat6)(\hat{1}\hat{2}\hat{3}\hat{4}\hat{5})=-\hat1\hat2\hat3\hat5\hat4+\hat1\hat2\hat3\hat5\hat6-\hat1\hat2\hat4\hat5\hat6+\hat1\hat3\hat4\hat5\hat6+\hat1\hat2\hat4-\hat1\hat3\hat4, \]
and
\[(\hat{1}\hat{2}\hat{3}\hat{4}\hat{5}\hat6)^2=\hat1\hat2\hat3\hat4\hat5\hat6-\hat1\hat2\hat3\hat6+\hat1\hat2\hat4\hat6-\hat1\hat3\hat4\hat6.  \]

\section{Future Research}
  One of the most interesting issues that arises in quantum mechanics is the crucial role played by the ``imaginary" number $i=\sqrt{-1}$. Many papers have been written suggesting, sometimes even ``proving", that it is impossible to formulate QM without complex numbers. In Clifford's geometric algebras many geometric quantities arise which have square $-1$. 
  
  The most famous example is the geometric algebra $\G_3$, sometime referred to as {\it biquaternions}. The biquaternions are algebraically isomorphic to the famous complex {\it Pauli matrices} of quantum mechanics. In  $\G_3$ the imaginary number $i$ takes on the geometric significance of an oriented three dimensional volume element, called a unit {\it trivector}. The unit trivector $i:=e_1e_2e_3$ is in the center of $\G_3$ and has square $-1$, where
  $e_1,e_2,e_3$ are anti-commuting unit vectors with that $e_1^2=e_2^2=e_3^2=1$, \cite{Hsta,gs}. The geometric algebra $\G_3$ has found many application in mathematics and theoretical physics, for example \cite{BandG,sob2023oz}.
  
  However, we have a problem. The algebra
  ${\cal A}_3^+\equiv \G_{1,2} \ne \G_3$. The problem is solved by realizing that whereas
  $\G_{1,2}\ne \G_3$, $\G_{1,2} \widetilde = \G_3$, the algebras are {\it isomorphic}. The relationship between the algebras is nicely expressed by 
  \[   \G_3:=\R(e_1,e_2,e_3)\widetilde= \R(e_1,if_1,-if_2)\widetilde= \R(e_1,f_1,f_2)=:\G_{1,2}, \]
  where we have made the identifications 
  \[ i:=e_1f_1f_2,\  e_2:=if_1=e_1f_2, \  e_3=-if_2=e_1f_1\]
  so that 
  \[  i =e_1(if_1)(-if_2) =e_1e_2e_3, \]
  \cite{SNF,SobHAL2023,Sob2023,Sob2019}.
  
  This demonstrates that the {\it quadratic form} determining the {\it metric} of a geometric algebra is inseparably tied up with the property of {\it anti-commutativity}. In the geometric $\G_3$, the metric is determined by the anti-commutativity of $e_1, e_2, e_3$, whereas the metric of the geometric algebra $\G_{1,2}$ is determined by the anti-commutativity of the elements $e_1,f_1,f_2$.
  
  In Einstein's {\it Special Relativity}, this dichotomy reveals the underlying structure of spacetime itself. In Hestenes' {\it Space Time Algebra}, the {\it Dirac Algebra} of gamma matrices is the matrix equivalent of the geometric algebra $\G_{1,3}\equiv {\cal A}_4^+$, \cite{BandG,Hsta,1905}. For example, each {\it timelike} Minkowski vector $\gamma_0$, such as $\gamma_0:=a_1+a_2$, determines the {\it rest frame} of an {\it observer}, together with her geometric algebra of observables, {\it bivectors} in the even subalgebra $\G_{1,3}^+(\gamma_0)$ of $\G_{1,3}$,
  \[ \G_3(\gamma_0):=\R(\be_1,\be_2,\be_3) \widetilde=\R(\gamma_1\gamma_0,\gamma_2\gamma_0,\gamma_3\gamma_0)\widetilde =:\G_{1,3}^+(\gamma_0), \]
  with the identification
  \[  \be_1:=f_1e_1, \ \be_2:=f_2 e_1, \ \be_3=f_3e_1,    \]  
  and where 
  \[\gamma_0:=e_1,\ \gamma_1:=f_1,\ \gamma_2:=f_2,
\ \gamma_3:=f_3,\]
\cite{Sob2019,st-vec-anal1981,gs}.

There are two very interesting representations of the symmetric group ${\cal S}_{n+1}$ that can be found in ${\cal A}_{n+1}^+$, \cite[Ch.12]{SNF}, \cite[Ch.5]{Sob2019}. It seems natural to identify a {\it qubit} as a pair of compatible null vectors; the {\it up state} of the qubit their {\it sum}, and the {\it down state} being their {\it difference}. The maximal quantum element of
order $n+1$ in the canonical basis of ${\cal A}_{n+1}^+$ is $\widehat{N}:=\hat 1 \cdots \hat j \cdots \hat k \cdots \widehat{n+1}$, where integers $1 \le j < k \le n+1$. The {\it two cycle} $(jk)$, which interchanges the quantum integers $\hat j$ and $\hat k$ in $\widehat N$, is specified by
\beq (jk)\widehat N :=- (\hat j-\hat k)\widehat N(\hat j - \hat k)=\hat 1 \cdots \hat k \cdots \hat j \cdots \widehat{n+1},  \label{permjk} \eeq
while leaving the positions of the other quantum integers in $\widehat N$ unchanged. 
It is seen that the two cycle is an {\it active Lorentz transformation}, represented by the {\it double covering} of the {\it Lorentz group} in $\G_{1,n}\equiv {\cal A}_{n+1}^+$, acting on the maximal quantum element $\widehat N$. 

The second representation of the symmetric group in ${\cal A}_{n+1}^+$ is specified by {\it triplets} of distinct positive quantum integers $\hat i, \hat j, \hat k \in {\cal A}_{n+1}^+$, acting on {\it quantum sums} of {\it pairs} of these quantum integers,
\beq (\hat i- \hat j)(\hat j + \hat k)(\hat i-\hat j)= \hat i+\hat k . \label{permsumgp} \eeq
The {\it passive Lorenz transformation} double covering takes the sum $\hat j+\hat k$ into the sum $\hat i + \hat k$. For ${\cal A}_3^+$, we have the three distinct pairs 
\[ \hat 1+\hat 2 \leftrightarrow 3, \ \hat 2 +\hat 3 \leftrightarrow 1,  \ \hat 3 + \hat 1 \leftrightarrow 2. \] 
The three $2$-cycles are specified by
\[  (\hat1-\hat2)(\hat 2+\hat 3)(\hat 1-\hat2)=(\hat 1+ \hat 3)  \]
\[  (\hat2-\hat3)(\hat 1+\hat 3)(\hat 2-\hat 3)=(\hat 1+ \hat 2)  \]
and
\[  (\hat3-\hat1)(\hat 1+\hat 2)(\hat 1-\hat 2)=(\hat 2+ \hat 3).  \]

In view of (\ref{permjk}) and (\ref{permsumgp}),
it seems natural to identify a {\it qubit} with  a pair of compatible null vectors; the {\it up state} being their {\it sum}, and the {\it down state} being their {\it difference}. It also appears that double covering Lorentz transformations can be used in defining the various logical gates \cite{TimD2001,TandAll2002}. Because of the differences in my approach, and that of Professor Kauffman, I have chosen the title of my paper to be {\it Itinerant Quantum Integers: The Language of Quantum Computing}, rather than using Professor Kauffman's {\it Iterants, and the Dirac Equation}, \cite{Kauf-Iterants2019}.  

Finally, it is important to remark that whereas all Clifford geometric algebras $\G_{p,q}$ are isomorphic to {\it real, complex}, or {\it split matrix algebras}, it is equally true that all geometric algebras $\G_{p,q}$ are isomorphic to
real, complex, or split quantum algebras ${\cal A}_{n+1}^+$, where $n+1=p+q$, \cite[p.217]{PL2001}, \cite[p.74]{Sob2019}, \cite{SobPeriodic2002}. The author believes that this preliminary work will find many applications in theoretical physics, representation theory, and many other areas of science and engineering. Perhaps, most importantly, it will become the natural language of quantum computers. The word {\it itinerant} means "wondering", perhaps this word captures the spirit of quantum integers and the mysterious nature of {\it entanglement} which makes quantum computers possible \cite{AntonJune14}.  

\section*{Acknowlegment}
The {\it Zbigniew Oziewicz Seminar on Fundamental Problemes in Physics}, organized by Professors Jesus Cruz and William Page has played an important role in the development of this work in relation to Category Theory and the Theory of Graphs, \cite{FES-C,ZO2013}.

\end{document}